\documentclass[aps,prl,superscriptaddress,twocolumn,longbibliography]{revtex4-1}
\usepackage{bbm}
\usepackage{graphicx}% Include figure files
\usepackage{dcolumn}% Align table columns on decimal point
\usepackage{bm}% bold math
\usepackage{subfigure}
\usepackage{amsmath}
\usepackage{feynmf}
\usepackage{hyperref}

\usepackage{attachfile}

\usepackage{times}

\def\be{\begin{equation}} \def\ee{\end{equation}}
\def\bea{\begin{eqnarray}} \def\eea{\end{eqnarray}}

\def\bk{{\bf k}}

\def\be{{\bf e}}
\def\bd{{\bf d}}

\def\bG{{\bf G}}

\begin{document}

\title{Higher-order topological odd-parity superconductors}
\author{Zhongbo Yan}
\email{yanzhb5@mail.sysu.edu.cn}
\affiliation{ School of Physics, Sun Yat-Sen University, Guangzhou 510275, China}

%\date{\today}

\begin{abstract}

The topological property of a gapped odd-parity superconductor is jointly determined by
its pairing nodes and Fermi surfaces in normal state. We reveal that the contractibility of
Fermi surfaces without crossing any time-reversal invariant momentum and the presence of
nontrivial Berry phase on  Fermi surfaces are two key
conditions for the realization of higher-order topological odd-parity superconductors (HOTOPSCs).
When the normal state is a normal metal, we reveal the necessity of
removable Dirac pairing nodes and provide a general and simple principle to realize
HOTOPSCs. Our findings can not only be applied to analyze the topological property
of odd-parity superconductors, but also be used as a guiding principle to design new platforms of
higher-order topological superconductors, as well as higher-order topological insulators
owing to their direct analogy in Hamiltonian description.

%In both two dimensions
%and three dimensions, we provide novel minimal models to demonstrate the physics.

%The combination of odd parity pairing and inversion
%symmetry requires an even number of disconnected Fermi surfaces in the normal state
%to avoid chiral edge states which are not compatible with Majorana corner modes.
%We find when there exist equal number of pairing nodes with opposite windings
%between (for Fermi surfaces concentric about time-reversal invariant momentum) or within
%(for Fermi surfaces not around any time-reversal invariant momentum) the Fermi surfaces, Majorana corner
%modes can emerge.

%By constructing novel minimal models
%of higher-order topological odd-parity superconductors (HOTOPSCs) in both two dimensions
%and three dimensions, we reveal that the presence of removable Dirac pairing nodes (RDPNs)
%at generic momentum is crucial to realize HOTOPSCs
%if, without superconductivity, the system corresponds to a normal metal. Imposing
%the inversion symmetry constraint,  we find the requirement on Fermi surface is that
%it is noncontractile (inevitably involving the closure of bulk gap when the Fermi surface
%continuously contracts  to a point) when RDPNs exist, but contractile after
%all RDPNs are removed.

\end{abstract}

%\pacs{71.10.Fd, 03.75.Ss, 74.25.Ha }
\maketitle

A defining characteristic of topological phases is their bulk-boundary correspondence,
namely, a topologically nontrivial  bulk will  manifest itself through
the boundary modes\cite{Chiu2016RMP}. Recently, higher-order topological  insulators (HOTIs) and
superconductors (HOTSCs) have attracted considerable interest owing to
the emergence of unconventional bulk-boundary correspondence\cite{benalcazar2017quantized,Song2017higher,Langbehn2017hosc,Benalcazar2017prb,Schindler2018HOTIa,ezawa2018higher,Rasmussen2018HOSPT,You2018hospt,
Khalaf2018hosc,Geier2018hosc,Franca2018HOTI,Roy2019HOTI,Trifunovic2019HOTI,Ahn2018HOTI,Kudo2019HOTMI}. As is known,
the boundary modes of conventional topological insulators (TIs) and topological superconductor (TSCs) are located at their one-dimensional lower boundaries\cite{hasan2010,qi2011},
however, for an $n$-th order TI or TSC with $n\geq2$, its boundary modes are located at its $n$-dimensional lower boundaries
(accordingly, conventional TIs and TSCs are also dubbed first-order TIs and TSCs, respectively).
In two dimensions ($2D$) and three dimensions ($3D$), such boundary modes are commonly dubbed corner modes
or hinge modes, and have been predicated to exist in quite a few materials\cite{yue2019symmetry,Wang2018XTe,Xu2019HOTI,Sheng2019SOTI,Lee2019HOTI,Chen2019HOTI} and
observed in a series of platforms, including photonic crystals\cite{noh2018topological,Chen2019photonic,Xie2108photonic,Hassan2019corner}, microwave resonators\cite{peterson2018quantized},
circuit arrays\cite{imhof2018corner}, phononic crystals\cite{serra2018observation,xue2019acoustic,zhang2019second},
bismuth\cite{schindler2018HOTI},  and iron-based superconductors\cite{Gray2019helical}. More recently, these concepts
have further been extended to cold atom systems\cite{Wu2019hosc,Zeng2019hosc}, Floquet systems\cite{Huang2018HOFTI,Peng2018HOFTI,Bomantara2019HOFTI,Seshadri2019FHOTI,Rodriguez-Vega2018HOFTI,Nag2019HOFTI,Hu2019HOFTI,Plekhanov2019FHOTSC}, as well as non-Hermitian systems\cite{Liu2019SOTI,Edvardsson2019,Lee2018NHHOTI,Ezawa2019NHHOTI,Luo2019NHHOTI,Zhang2019NHHOTI}.

The boundary modes of HOTSCs  are of particular interest for their potential application in topological quantum computation\cite{You2018hosc,Bomantara2019braiding,Pahomi2019braiding}.
Thus far, a general approach to realize HOTSCs is ``order transition''\cite{Shapourian2018sotsc,Zhu2018hosc,Yan2018hosc,Wang2018hosc,Wang2018hosc2,Liu2018hosc,Hsu2018,Pan2018SOTSC,
Bultinck2019HOSC,Yang2019hinge,Volpez2019SOTSC,Zhang2019hinge,Kheirkhah2019SOTSC,Wu2019swave,Laubscher2019FHOSC,Zhang2019HOTSC},  that is, by breaking
certain appropriate symmetries, the one-dimensional lower boundary modes will be gapped out in a nontrivial way, and accordingly,
the first-order topological phase is transited to a higher-order topological phase. According to this  approach,
if the starting first-order topological phase is an odd-parity superconductor, to gap out its one-dimensional
lower boundary modes, one has to introduce terms of even parity to break certain symmetries\cite{Wu2019hosc,Zhu2018hosc,Wang2018hosc2}.  This, while suggesting
that superconductors with appropriate mixed-parity pairings are candidates of HOTSCs\cite{Wu2019hosc,Wang2018hosc},
does not mean that odd-parity pairing only can not realize intrinsic HOTSCs. In fact, the authors in ref.\cite{Wang2018hosc}
have demonstrated that a Dirac semimetal with chiral $p$-wave pairing provides a realization of
second-order TSC in $2D$. Nevertheless, a general theory of intrinsic higher-order topological odd-parity superconductors(HOTOPSCs)
is still lacking. In particular, we notice that when the normal state is a featureless normal metal, what kind of
pairing and Fermi surface structure can realize intrinsic HOTOPSCs has not been explored.

As the topological property of an odd parity superconductor is jointly determined by
its pairing nodes and Fermi surfaces in normal state\cite{sato2010odd,fu2010odd}, in this work, we investigate the general conditions on
pairing nodes and Fermi surfaces for the realization of HOTOPSCs. Our study
reveals that there are two key conditions for the realization of HOTOPSCs.
One is that the Fermi surfaces can continuously contract to a point without
crossing any time-reversal invariant (TRI) momentum, and the other is
the presence of nontrivial Berry phase on the Fermi surfaces. Importantly,
when the normal state is a normal metal, we reveal the necessity of
removable Dirac pairing nodes (RDPNs) and provide a general and simple principle
to realize HOTOPSCs.

{\it General theory.---} Given $H=\sum_{k}\Psi_{\bk}^{\dag}H(\bk)\Psi_{\bk}$ with
$\Psi_{\bk}=(c_{\bk}, c_{-\bk}^{\dag})^{T}$, the topological property of a superconductor
is encoded in $H(\bk)$ whose general form is given by
\begin{eqnarray}
H(\bk)=\left(
         \begin{array}{cc}
           \varepsilon(\bk) & \Delta(\bk) \\
           \Delta^{\dag}(\bk) & -\varepsilon(\bk) \\
         \end{array}
       \right),\label{general}
\end{eqnarray}
where $\varepsilon(\bk)$ describes the normal state and
$\Delta(\bk)$ represents the pairing order parameter.
In this work, we focus on inversion symmetric normal states and
odd-parity pairings which satisfy  $\Delta(\bk)=-\Delta(-\bk)$. Apparently,
$\Delta(\bk)$  always vanishes at TRI
momenta in the Brillouin zone, i.e., momenta satisfy $\bk=-\bk+m\bG$ with $\bG$ the reciprocal lattice vector
and $m=0$ or $1$. This means that  the pairing nodes at TRI
momenta (TRIPNs)  are unmovable and unremovable. When the normal state is a normal metal,
the TRIPNs of a gapped odd-parity superconductor are of Dirac point nature,
so when a Fermi surface encloses one TRIPN, it has a nontrivial Berry phase as
the pairing order parameter shows a nonzero integer times of winding on it. The presence of nontrivial Berry phase
on Fermi surfaces is the origin of nontrivial topology.

In $2D$ and $3D$, it has been demonstrated that if the number of Fermi surfaces
enclosing TRI momentum (for TRI systems, the number does not take into
account the Kramers degeneracy) is odd, a gapped odd-parity superconductor is a first-order TSC\cite{sato2010odd,fu2010odd}.
This implies that to guarantee the first order
topological property to be trivial, the Fermi surfaces must be contractile in the sense
that it can continuously contract
to a point without crossing any TRI momentum. Noteworthily, however, this does not
mean that the Fermi surfaces can directly contract to a point without closing
the bulk gap as there may exist other Dirac pairing nodes at generic momentum.
In fact,  as HOTOPSCs are essentially distinct to trivial superconductors in topology,
one can conjecture that to realize HOTOPSCs,  the presence of nontrivial Berry phase
on  Fermi surfaces should be  necessary. There are two ways to achieve this,
one is that the Fermi surfaces enclose Dirac pairing nodes away from TRI momentum,
and the other is that the underlying normal state is a topological semimetal for which
the band touchings themselves will contribute nontrivial Berry phase, like in ref.\cite{Wang2018hosc}.
Thus, if the normal state is a normal metal, the existence of Dirac pairing nodes away from TRI momentum
should be necessary for realizing HOTOPSCs.

\begin{figure}
\subfigure{\includegraphics[width=4cm, height=4cm]{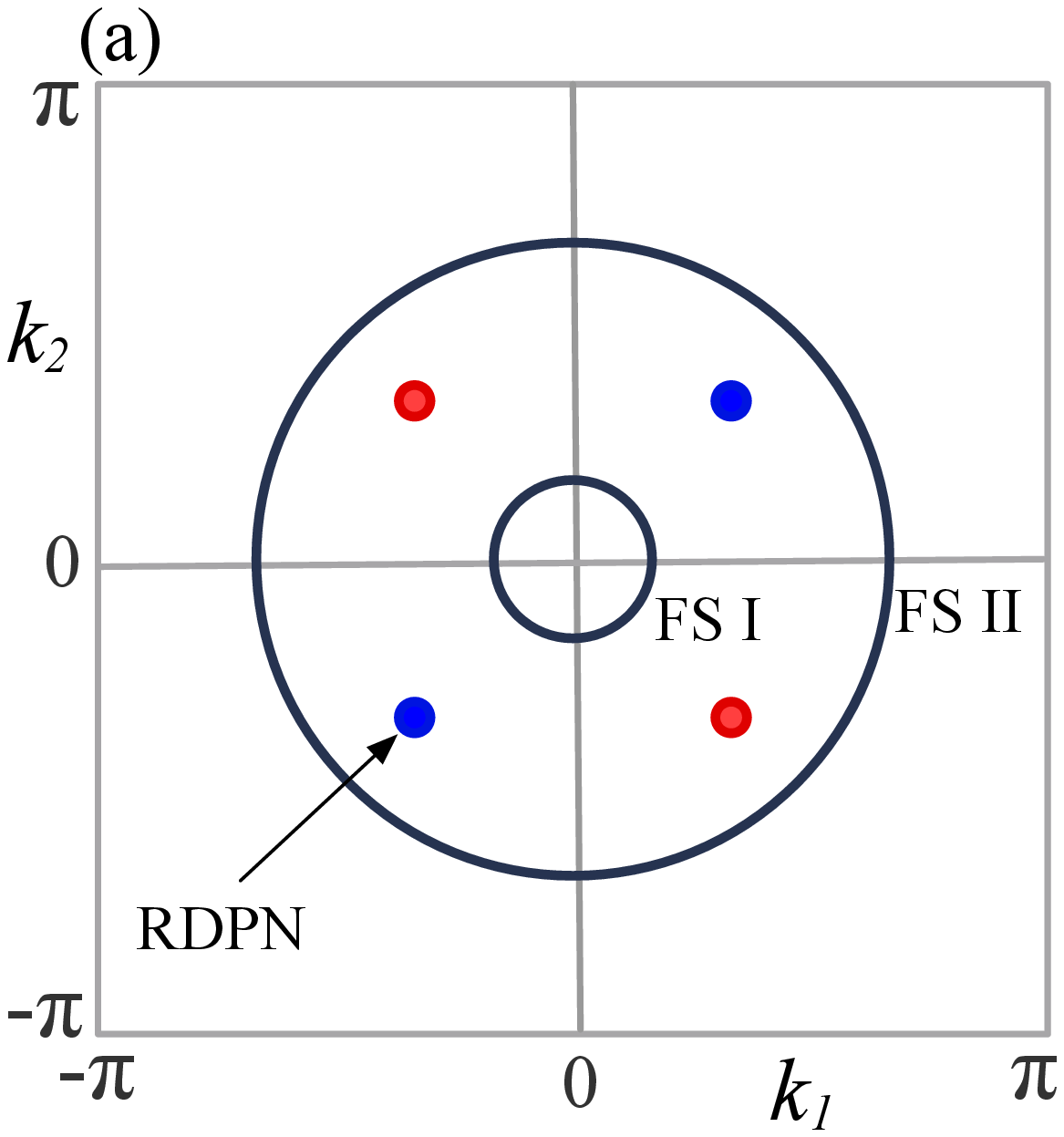}}
\subfigure{\includegraphics[width=4cm, height=4cm]{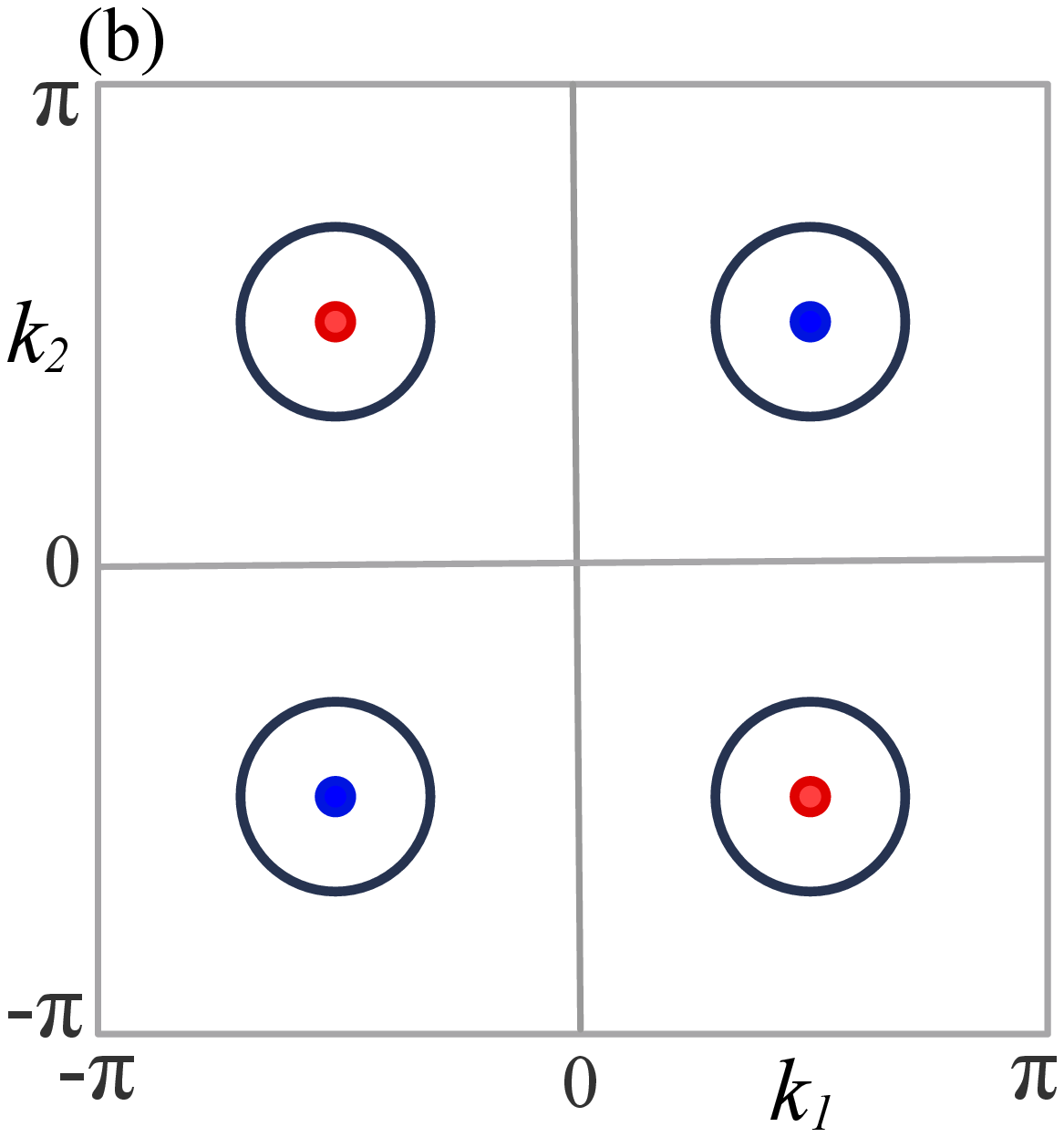}}
\caption{ Two  representative configurations of Fermi surfaces and RDPNs that realize SOTOPSCs in $2D$.
The circles in black represent the Fermi surfaces, and the dots with different color represent RDPNs
with opposite winding number.}  \label{sketch}
\end{figure}

{\it Second-order topological odd-parity superconductors (SOTOPSCs) in $2D$.---}
In $2D$,
a novel class of  models with odd-parity pairing and trivial Chern number (so trivial first-order topological
property) can be constructed by a novel approach called Hopf map\cite{Yan2018hopf}.
According to this approach, we let
\begin{eqnarray}
H(\bk)=d_{1}(\bk)\tau_{1}+d_{2}(\bk)\tau_{2}+d_{3}(\bk)\tau_{3}\label{minimal}
\end{eqnarray}
with $d_{i}(\bk)=z(\bk)^{\dag}\tau_{i}z(\bk)$, where
$z_{1}(\bk)=f_{1}(\bk)+if_{2}(\bk)$, $z_{2}(\bk)=g_{1}(\bk)+ig_{2}(\bk)$
and $\tau_{1,2,3}$ are Pauli matrices in particle-hole space. To describe
an odd-parity superconductor, we let $f_{i=1,2}(\bk)$ be real and even functions of
momentum, i.e., $f_{i}(\bk)=f_{i}(-\bk)$, and let $g_{i=1,2}(\bk)$ be real and odd functions of
momentum, i.e., $g_{i}(\bk)=-g_{i}(-\bk)$. It is noteworthy that this choice of $f_{i}$ and $g_{i}$ is distinct to
the conventional Hopf map\cite{Moore2008hopf,Deng2013hopf,Yan2017link,Ezawa2017hopf}. A comparison of Eq.(\ref{general}) and
Eq.(\ref{minimal}) reveals
\begin{eqnarray}
\Delta(\bk)&=&d_{1}(\bk)-id_{2}(\bk)=2[f_{1}(\bk)+if_{2}(\bk)][g_{1}(\bk)-ig_{2}(\bk)],\nonumber\\
\varepsilon(\bk)&=&d_{3}(\bk)=f_{1}^{2}(\bk)+f_{2}^{2}(\bk)-g_{1}^{2}(\bk)-g_{2}^{2}(\bk).\label{expression}
\end{eqnarray}
Before giving concrete expressions to $f_{i}(\bk)$ and $g_{i}(\bk)$,
we make a general discussion about the Hamiltonian above. Clearly,
the different parity of $f_{i}(\bk)$ and $g_{i}(\bk)$ guarantees
$\Delta(\bk)=-\Delta(-\bk)$, confirming that it describes an odd-parity  superconductor.
Moreover, according to the expression of $\Delta(\bk)$ in
Eq.(\ref{expression}),  one can  find that Dirac pairing nodes will show up at generic momentum
when $f_{1}(\bk)=0$ and $f_{2}(\bk)=0$ can simultaneously be satisfied.
In contrast to TRIMPNs, such Dirac pairing nodes are removable. Focusing on
the Fermi surface  determined by  $\varepsilon(\bk)=0$,
one can further find  that the number of disconnected Fermi surfaces must be even
and the removable Dirac pairing nodes (RDPNs), if they exist,
are  located within the disconnected  Fermi surfaces
or between two near neighbour disconnected Fermi surfaces (see Fig.\ref{sketch}
for a graphic illustration), which guarantees that the Fermi surfaces can not continuously
contract to a point without closing the bulk gap. When
each disconnected Fermi surface encloses an odd number of Dirac pairing nodes,
the presence of nontrivial Berry phase on Fermi surfaces will also be satisfied.

While we have infinite choices on $f_{i}(\bk)$ and $g_{i}(\bk)$,
in this work we let
\begin{eqnarray}
f_{i}(\bk)&=&f_{i}(k_{i})=(\cos k_{i}+\lambda_{i}), \nonumber\\
g_{i}(\bk)&=&g_{i}(k_{i})=\sin k_{i}.\label{convention}
\end{eqnarray}
Accordingly, we have
$d_{1}(\bk)=2\sum_{i=1,2}(\cos k_{i}+\lambda_{i})\sin k_{i}$,
$d_{2}(\bk)=2(\cos k_{1}+\lambda_{1})\sin k_{2}-2(\cos k_{2}+\lambda_{2})\sin k_{1}$ and
$d_{3}(\bk)=\sum_{i=1,2}(\cos k_{i}+\lambda_{i})^{2}-\sin^{2}k_{i}$. Meanwhile,
the energy spectra are
\begin{eqnarray}
E(\bk)=\pm\sum_{i=1,2}[(\cos k_{i}+\lambda_{i})^{2}+\sin^{2} k_{i}],\label{spectra}
\end{eqnarray}
one can find that the bulk gap vanishes only when $|\lambda_{1}|=|\lambda_{2}|=1$.
As an even-parity term can be taken as a Dirac mass, the presence of two Dirac masses
in Eq.(\ref{spectra}) guarantees the  first-order topological property to be trivial.

According to Eq.(\ref{convention}), when
$|\lambda_{1,2}|<1$,  the RDPNs  are located at
$\bk=(\pm Q_{1}, \pm Q_{2})$ with $Q_{1,2}=\pi-\arccos\lambda_{1,2}$, and one can find
that the configuration of Fermi surfaces and RDPNs belongs to the type shown in Fig.\ref{sketch}(a).
When $|\lambda_{1}|=1$ or $|\lambda_{2}|=1$, the RDPNs coincide in pairs and annihilate.
Once  $|\lambda_{1}|>1$ or $|\lambda_{2}|>1$, they are removed.  For each pairing node,
we can assign a winding number to characterize its topological property,
\begin{eqnarray}
w_{n}&=&\frac{1}{2\pi}\oint_{c} dk\frac{d_{1}\partial_{k}d_{2}-d_{2}\partial_{k}d_{1}}{d_{1}^{2}+d_{2}^{2}}\nonumber\\
&=&\frac{1}{2\pi}\oint_{c} dk[\frac{g_{1}\partial_{k}g_{2}-g_{2}\partial_{k}g_{1}}{g_{1}^{2}+g_{2}^{2}}
-\frac{f_{1}\partial_{k}f_{2}-f_{2}\partial_{k}f_{1}}{f_{1}^{2}+f_{2}^{2}}],
\end{eqnarray}
where $C$ denotes a closed contour enclosing only one pairing node. As
$f_{i}$ and $g_{i}$ decouple from each other, this indicates
that the creation or annihilation of RDPNs does not affect the
topological property of TRIMPNs. Such a property is in fact also crucial for
the  realization of SOTOPSCs. As a counter example, if we keep the form of $\varepsilon(\bk)$
and  let
$\Delta(\bk)=f_{2}g_{1}+if_{1}g_{2}$, while the locations of pairing nodes
are same, now all RDPNs have same winding number, consequently
the creation or annihilation of RDPNs will directly change the topological
property of TRIMPNs since the net winding number of all Dirac pairing nodes should
be zero. For this case, when $|\lambda_{1,2}|<1$,
the Hamiltonian in fact realizes  a first-order TSC with large Chern number,
instead of a SOTOPSC we want.

\begin{figure}
\subfigure{\includegraphics[width=4cm, height=4cm]{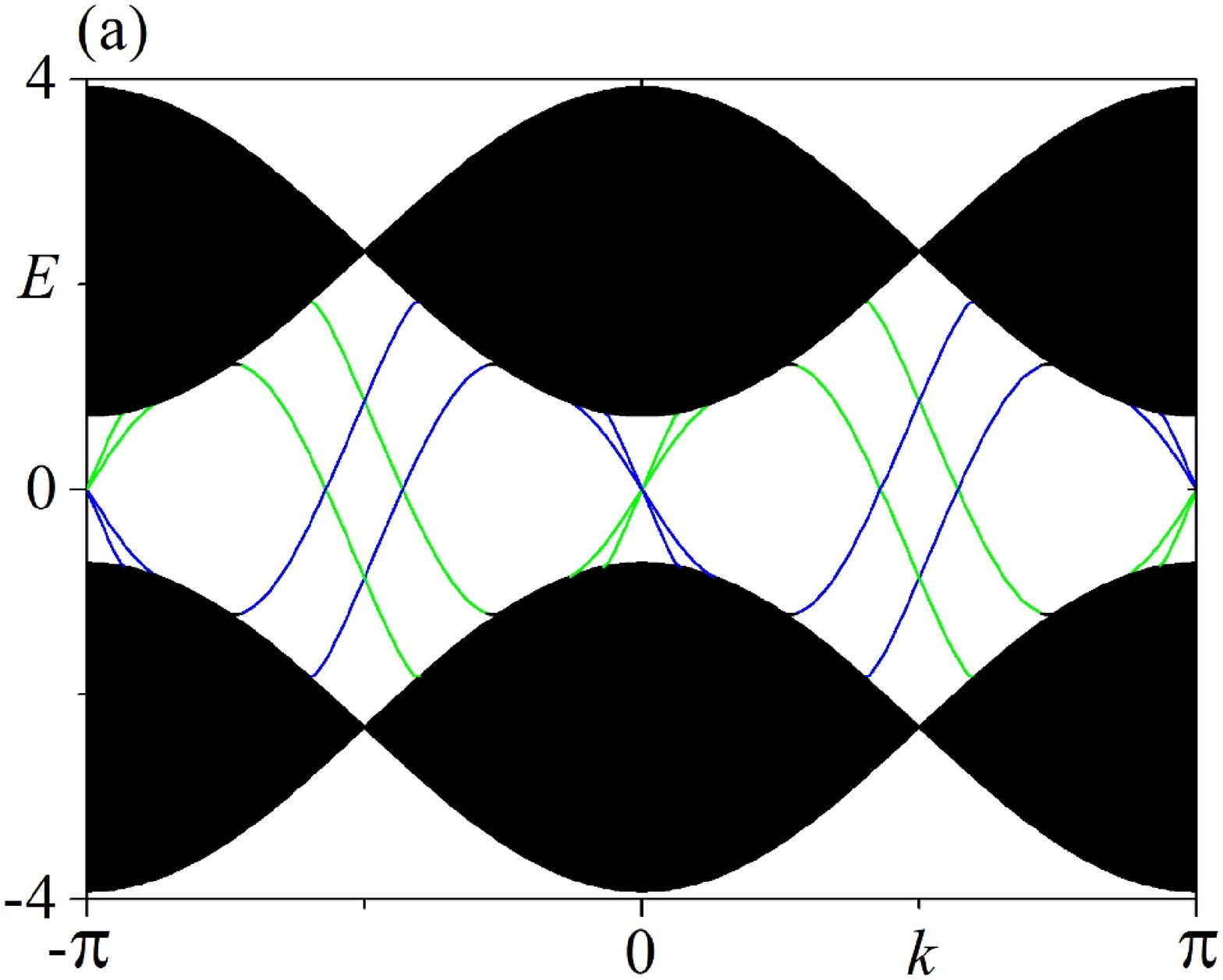}}
\subfigure{\includegraphics[width=4cm, height=4cm]{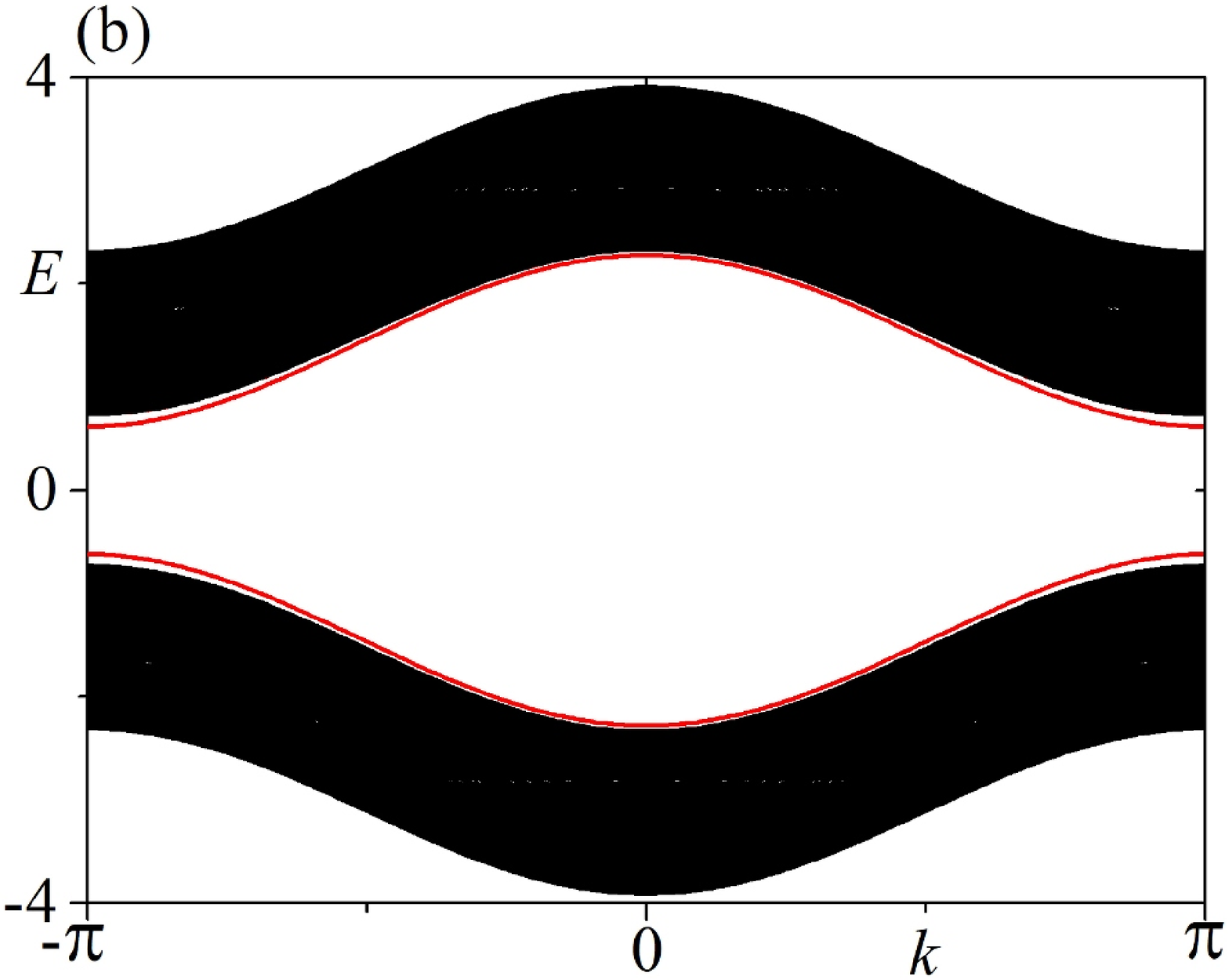}}
\subfigure{\includegraphics[width=4cm, height=4cm]{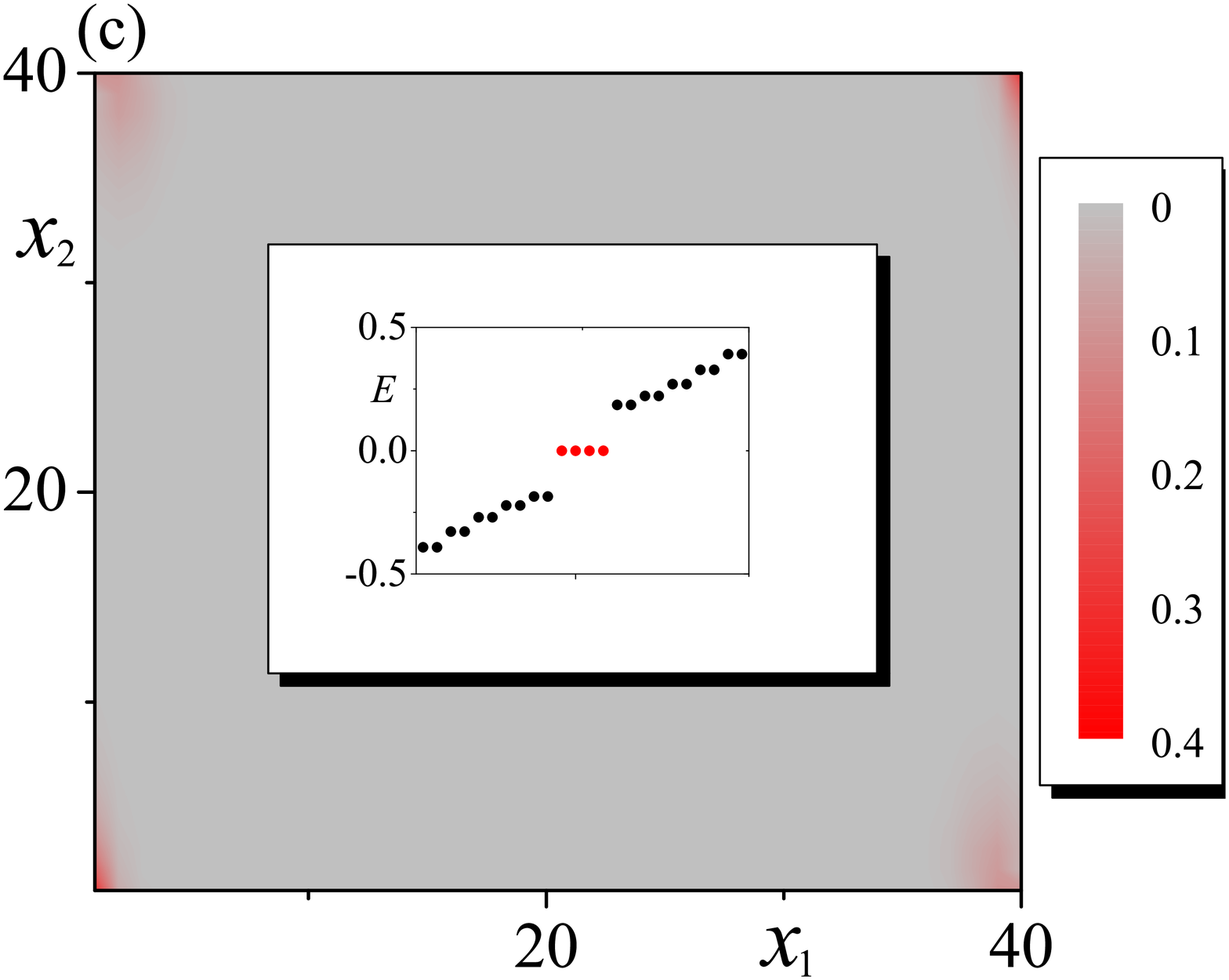}}
\subfigure{\includegraphics[width=4cm, height=4cm]{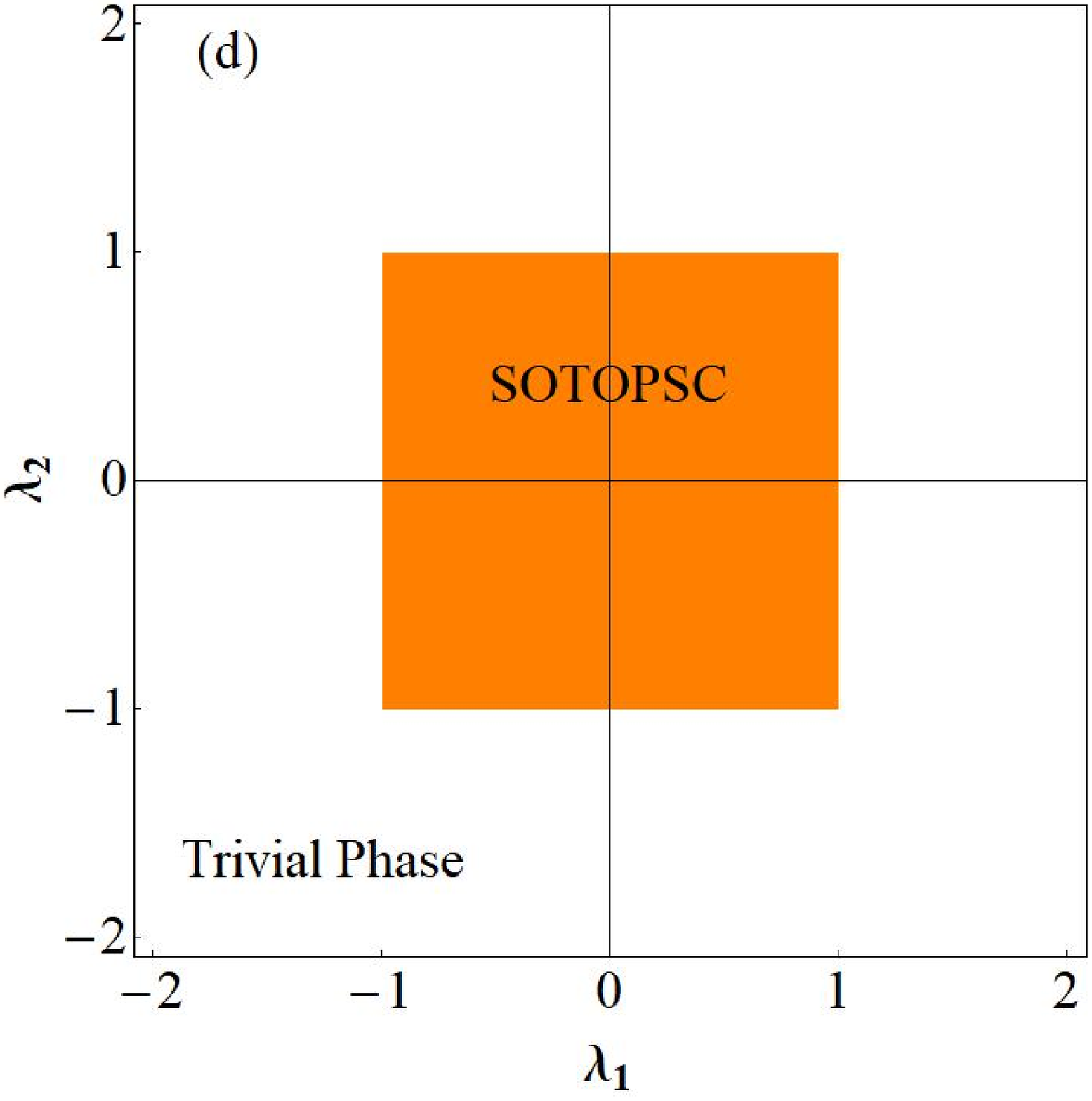}}
\caption{ Common parameters in (a)(b)(c): $\lambda_{1}=\lambda_{2}=0.4$.
(a) Energy spectra for a ribbon geometry with open boundary condition in the
$\hat{x}_{1}+\hat{x}_{2}$ direction along which the lattice size is $L_{\hat{x}_{1}+\hat{x}_{2}}=200$.
The gapless modes in blue and green colors are localized at the upper and lower edge, respectively.
(b) Energy spectra for a ribbon geometry with open boundary condition in
the $\hat{x}_{1}$ direction, $L_{1}=200$. The spectra in red represent gapped edge modes.
(c) MCMs in a finite-size system, $L_{1}=L_{2}=40$.  (d) Phase diagram.}  \label{second}
\end{figure}

To see that the Hamiltonian indeed realizes a SOTOPSC when RDPNs exist, i.e., $|\lambda_{1,2}|<1$,
let us focus on the special case with $\lambda_{1}=\lambda_{2}$
for an intuitive understanding.  When $\lambda_{1}=\lambda_{2}=\lambda$, there are
two special lines in the Brillouin zone, $k_{1}=k_{2}$ and $k_{1}=-k_{2}$,
on which chiral symmetry is preserved and thus a winding number can be defined.
On the $k_{1}=k_{2}$ line (the case with $k_{1}=-k_{2}$
can similarly be analyzed),  the Hamiltonian
reduces to
\begin{eqnarray}
H_{\rm R}(q)=d_{1}(q)\tau_{1}+d_{3}(q)\tau_{3},
\end{eqnarray}
where $q$ represents the momentum along the line
$k_{1}=k_{2}$, $d_{1}(q)=4(\cos q+\lambda_{1})\sin q$, and
$d_{3}(q)=2(\cos q+\lambda_{1})^{2}-2\sin^{2} q$.
The winding number characterizing the topological property of
$H_{\rm R}(q)$ is given by
\begin{eqnarray}
w_{\rm R}=\frac{1}{2\pi}\int_{-\pi}^{\pi} dq\frac{d_{3}\partial_{q}d_{1}-d_{1}\partial_{q}d_{3}}{d_{1}^{2}+d_{3}^{2}}
=\left\{
     \begin{array}{cc}
       2, & |\lambda|<1, \\
       0, & |\lambda|<1. \\
     \end{array}
   \right.\label{winding}
\end{eqnarray}
The result indicates when $\lambda_{1}=\lambda_{2}$ and
$|\lambda_{1,2}|<1$,
the Hamiltonian describes a weak TSC.
Accordingly, If the system is of a ribbon geometry and
open boundary condition is taken in the $\hat{x}_{1}+\hat{x}_{2}$ (or $\hat{x}_{1}-\hat{x}_{2}$)
direction, then gapless modes will show up on the edges.
On each edge, the number of left-moving modes and right-moving modes
must be equal as the bulk Chern number is zero.  As shown in
Fig.\ref{second}(a), when $|\lambda|<1$, each edge indeed harbors four
left-moving modes and four right-moving modes, confirming the expectation.
It is noteworthy that the number of gapless modes is four times the winding number
given in  Eq.(\ref{winding}), which is because the $d_{2}$ term has four zeroes
along the line $k_{1}=-k_{2}$.
As a comparison, if open boundary condition
is not along the $\hat{x}_{1}+\hat{x}_{2}$ (or $\hat{x}_{1}-\hat{x}_{2}$)
direction, one can expect the absence of gapless edge states. In Fig.\ref{second}(b),
the result for open boundary condition in the $\hat{x}_{1}$ direction
is presented, which clearly demonstrates the absence of gapless edge states
within the energy gap.

The defining characteristic of a SOTOPSC in $2D$ is the presence of
Majorana corner modes (MCMs)\cite{Zhu2018hosc,Yan2018hosc,Wang2018hosc,Wang2018hosc2}.
By choosing open boundary condition in both the $\hat{x}_{1}$ and $\hat{x}_{2}$ directions,
we indeed find when $|\lambda|<1$,  each corner of the finite-size system harbors one Majorana zero mode (MZM),
as shown in Fig.\ref{second}(c). Here the presence
of MCMs can intuitively be understood by noting that the $d_{2}$ term
is odd under the
mirror reflection about the line $k_{1}=k_{2}$. From a low-energy perspective, this
implies that the Dirac mass gapping out the gapless edge modes will have opposite sign if the respective edges
are located at different sides of the $\hat{x}_{1}+\hat{x}_{2}$ (or $\hat{x}_{1}-\hat{x}_{2}$) direction. As
a result, if the $\hat{x}_{1}+\hat{x}_{2}$ (or $\hat{x}_{1}-\hat{x}_{2}$) direction places in
between a corner, the corner is a domain wall of Dirac mass and consequently harbors
a MZM. While such an intuitive picture relies on $\lambda_{1}=\lambda_{2}$,
the existence of MCMs does not rely on it.
Through detailed numerical calculation, we confirm that the MCMs
persist as long as $|\lambda_{1,2}|<1$, so the phase diagram in Fig.\ref{second}(d)\cite{supplemental}.

Before ending this part, we point out that the SOTOPSC follows a $Z_{2}$
classification because when two MZMs appear at the same corner,
there is no symmetry to protect them from coupling, and so splitting.
This implies when there exist many RDPNs and disconnected Fermi surfaces in the Brillouin zone,
if the structure of Fermi surfaces and pairing nodes can continuously evolve to the two
types of representative configurations given in Fig.\ref{sketch} without closing
the bulk gap, the system realizes a robust SOTOPSC.  Furthermore, it is worthy to point out
that to the best of our knowledge, Eq.(\ref{minimal}) is the first ``$\bd\cdot \tau$'' model
that realizes a second-order TSC (SOTSC) in $2D$. If putting two copies of the model together,  i.e.,
\begin{eqnarray}
H(\bk)=d_{1}(\bk)\tau_{1}s_{3}+d_{2}(\bk)\tau_{2}+d_{3}(\bk)\tau_{3}\label{TRI}
\end{eqnarray}
with $s_{1,2,3}$ the Pauli matrices in spin space, one also
obtains a minimal-model realization of TRI SOTSCs in $2D$.

{\it SOTOPSCs in $3D$.---} The scenario in two dimensions can naturally be generalized
to higher dimensions. To see this, we construct the following Hamiltonian,
\begin{eqnarray}
H(\bk)&=&\tilde{d}_{1}(\bk)\tau_{1}s_{1}+\tilde{d}_{2}(\bk)\tau_{1}s_{3}+\tilde{d}_{3}(\bk)\tau_{2}+\tilde{d}_{4}(\bk)\tau_{3},\,\,\label{third}
\end{eqnarray}
where $\tilde{d}_{1,2}(\bk)=d_{1,2}(\bk)$, $\tilde{d}_{3}(\bk)=\sin k_{3}$,
and $\tilde{d}_{4}(\bk)=d_{3}(\bk)-t(\cos k_{3}-1)$. The Hamiltonian describes
a three-dimensional TRI odd-parity superconductor, with $\tilde{d}_{1,2,3}(\bk)$ corresponding
to the pairings, and $\tilde{d}_{4}(\bk)$ characterizing the energy dispersion of the normal state.

For this Hamiltonian, RDPNs also exist only when $|\lambda_{1,2}|<1$. With the increase of dimension
to $3D$, the topological invariant characterizing Dirac pairing nodes needs to
be generalized as
\begin{eqnarray}
\nu_{n}=\frac{1}{4\pi}\oint_{S}dk^{2}\frac{\epsilon_{ijk} \tilde{d}_{i}\partial_{k_{\alpha}}
 \tilde{d}_{j} \partial_{k_{\beta}}  \tilde{d}_{k}}{(\tilde{d}_{1}^{2}+\tilde{d}_{2}^{2}+\tilde{d}_{3}^{2})^{3/2}}.
\end{eqnarray}
where $S$ denotes a closed surface enclosing one pairing node, $k_{\alpha}$ and $k_{\beta}$
are local coordinates characterizing $S$, and $\epsilon_{ijk}$ with $\{i,j,k\}=\{1,2,3\}$
is the Levi-Civita symbol. One can check that for this Hamiltonian, the creation
or annihilation of RDPNs also does not change the topological property of TRIMPNs, fulfilling
the requirement on RDNPs.

\begin{figure}
\subfigure{\includegraphics[width=4cm, height=4cm]{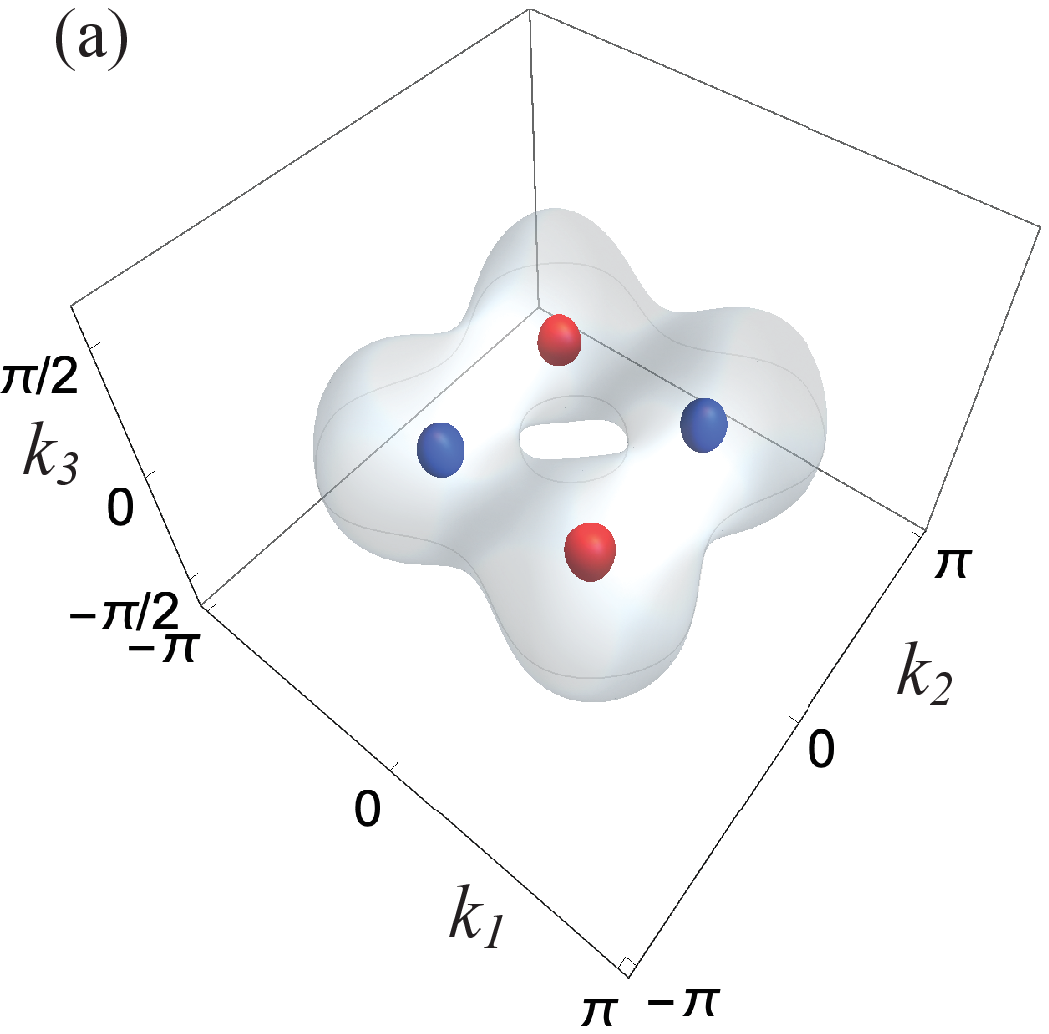}}
\subfigure{\includegraphics[width=4cm, height=4cm]{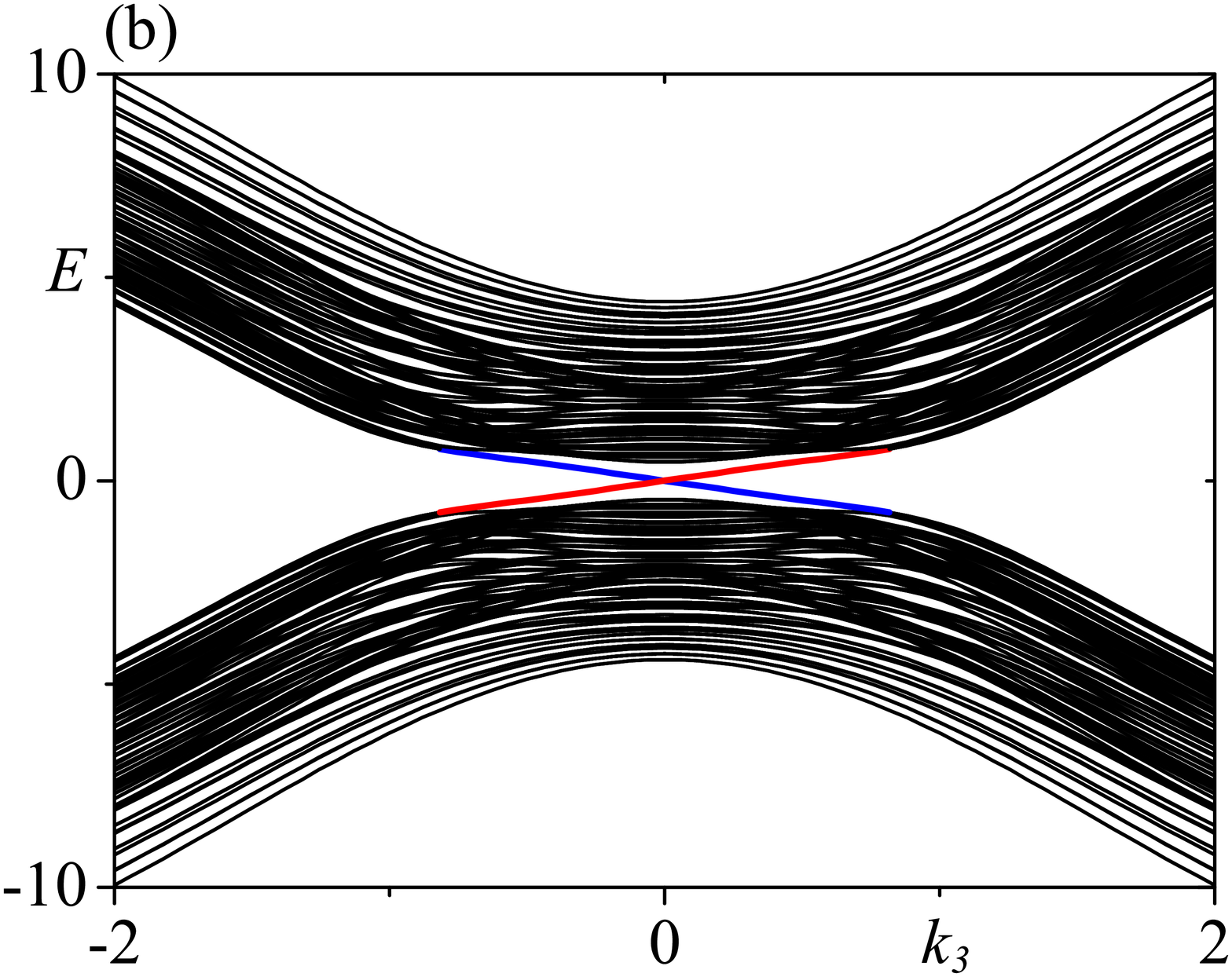}}
\subfigure{\includegraphics[width=4cm, height=3cm]{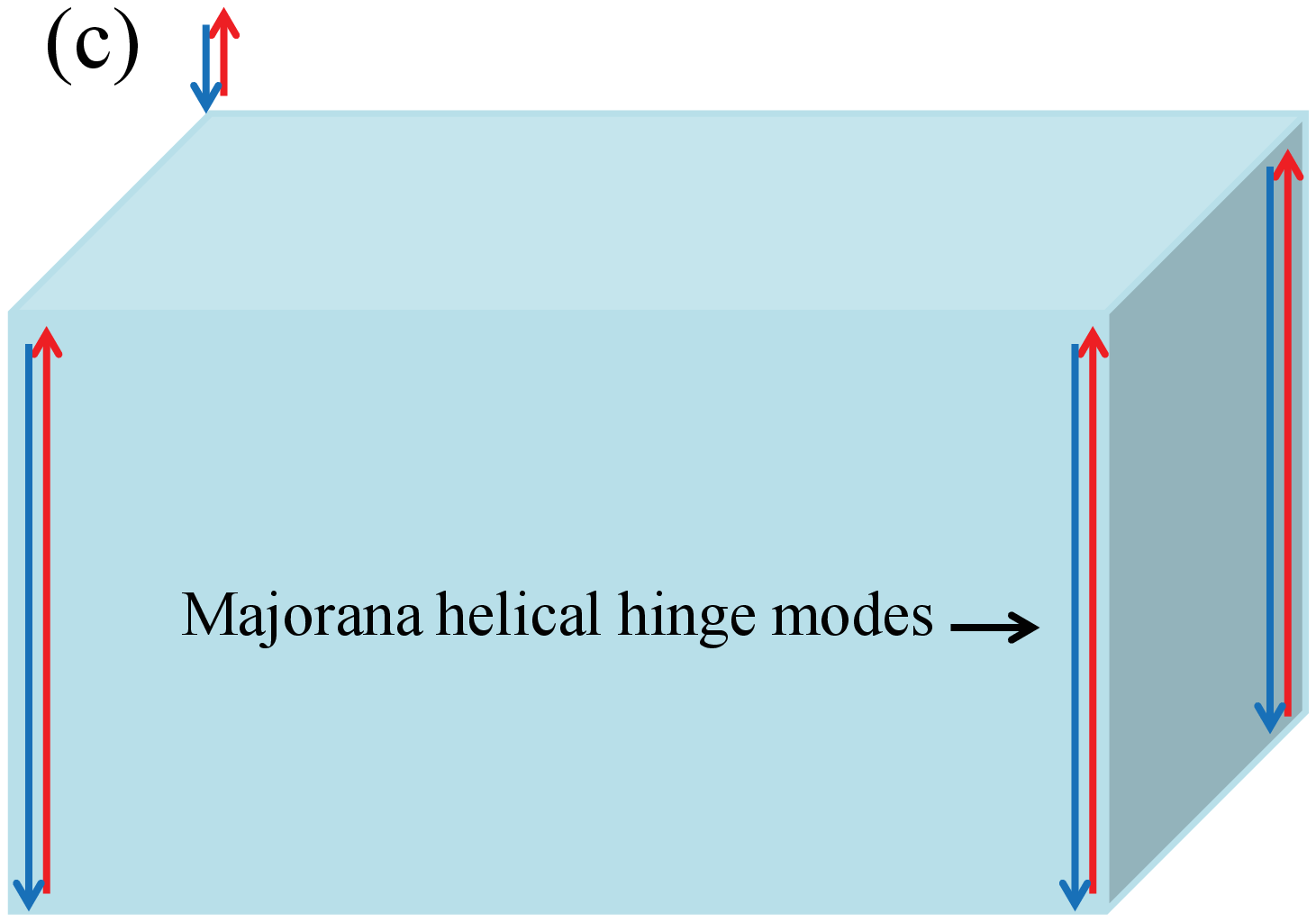}}
\subfigure{\includegraphics[width=4cm, height=3cm]{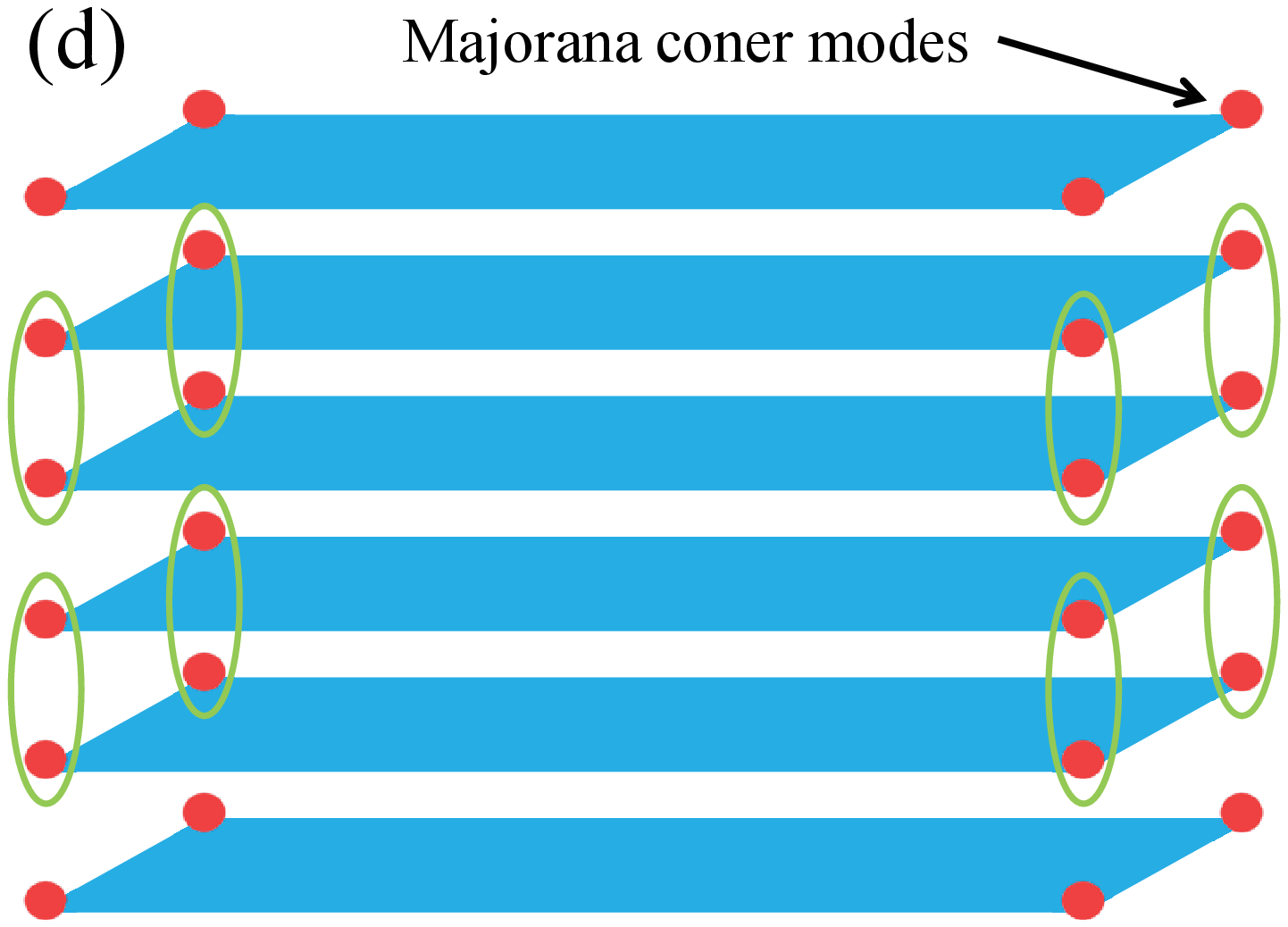}}
\caption{ (a) Fermi surface and RDPNs in the $k_{3}=0$ plane.
Parameters are $\lambda_{1}=\lambda_{2}=-0.5$, and $t=4$. The RDPNs (red and blue dots, the two colors represent
opposite topological charges) are located within the Fermi surface of torus
form. (b) Energy spectra for a sample with open boundary condition in both the $\hat{x}_{1}$ and
$\hat{x}_{2}$ direction, and periodic boundary condition in the $\hat{x}_{3}$ direction. The parameters are the same as in
(a), and $L_{1}=L_{2}=10$. The gapless  modes are of four-fold degeneracy. (c) A schematic illustration of the distribution of
the four branches of helical Majorana modes in (b). (d) A schematic illustration of the layer
construction of TOTOPSCs based on SOTOPSCs. }  \label{thinge}
\end{figure}

The Fermi surface is determined by  $\tilde{d}_{4}(\bk)=0$. One can find
when $t>t_{c}=1-(\lambda_{1}^{2}+\lambda_{2}^{2})/4$,
the Fermi surface only encloses the  RDPNs located at the $k_{3}=0$ plane (see Fig.\ref{thinge}(a)).
While the Fermi surface can continuously contract to a point without crossing any TRI momentum,
it can not continuously contract to a point without closing the bulk gap before the annihilation of RDPNs,
indicating that the Hamiltonian realizes a HOTOPSC when $t>t_{c}$ and $|\lambda_{1,2}|<1$.
It is noteworthy that while in the following we only consider $t>t_{c}$, the phases in the regime $t<t_{c}$ and $|\lambda_{1,2}|<1$
are also of great interest, e.g., a weak HOTOPSC will emerge when the Fermi surfaces enclose all RDPNs at both the
$k_{3}=0$ and $k_{3}=\pi$ planes\cite{supplemental}.

To confirm the realization of HOTOPSC when $t>t_{c}$ and $|\lambda_{1,2}|<1$,
we consider that the system takes open boundary condition
in both the $\hat{x}_{1}$ and  $\hat{x}_{2}$ directions, and periodic boundary condition in
the $\hat{x}_{3}$ direction. As shown in Fig.\ref{thinge}(b)(c), the numerical results reveal
that each hinge of the sample harbors a pair of Majorana helical modes, confirming
the realization of a TRI SOTOPSC in $3D$.

While it is apparent that this scenario can further be generalized to even higher dimensions, we notice
that the results in $2D$ and $3D$ strongly suggest that when the normal state is a normal metal,  only SOTOPSCs
can be realized. This limitation can be understood by noting the fact
that for normal state being a normal metal, the Fermi surface can have nontrivial Berry phase only when it encloses
 Dirac pairing nodes, but this goes back to the scenario above. Therefore, to realize
third-order topological odd-parity superconductors (TOTOPSCs), the underlying normal state needs to be a topological semimetal which itself
has some topological structure.

{\it TOTOPSCs in $3D$.---} A TOTOPSC in $3D$ can be realized by stacking
two dimensional SOTOPSCs layer by layer in a dimerized way, as illustrated in Fig.\ref{thinge}(d).
As an example, we construct the below Hamiltonian,
\begin{eqnarray}
H(\bk)&=&d_{1}(\bk)\tau_{1}s_{3}+d_{2}(\bk)\tau_{2}+d_{3}(\bk)\tau_{3}\sigma_{3}\nonumber\\
&&+(\cos k_{3}+\lambda_{3})\tau_{3}\sigma_{1}+\sin k_{3}\tau_{1}s_{1},\label{layer}
\end{eqnarray}
where $\sigma_{1,2,3}$ are Pauli matrices, e.g., in orbital space. One can see
that the first three terms realize the two dimensional SOTOPSC, while
the last two terms  realize a Kitaev chain
in the layer-stacking direction. The experience from Kitaev model tells us that
the situation presented in Fig.\ref{thinge}(d) corresponds to the limiting case
$\lambda_{3}=0$\cite{Kitaev2001unpaired}. For this special case,  the outer two layers decouple from
the inner layers, so MCMs will show up if
each layer realizes a SOTOPSC.  The experience from Kitaev model also
tells us that the model is within the same phase for $|\lambda_{3}|<1$\cite{Kitaev2001unpaired},
thus the Hamiltonian in Eq.(\ref{layer}) realizes a TOTOPSC
when $|\lambda_{1,2,3}|<1$\cite{supplemental}.

The normal state of the Hamiltonian in Eq.(\ref{layer}) is described by
$H_{N}(\bk)=d_{3}\sigma_{z}+(\cos k_{3}+\lambda_{3})\sigma_{x}$,
which turns out to be a nodal-line semimetal in the regime $|\lambda_{1,2,3}|<1$.
If weakly doping the normal state, each piece of Fermi surfaces is a thin torus enclosing
a nodal line. Along the poloidal direction, there is a global $\pi$-Berry phase\cite{fang2016nlsm}.
When $|\lambda_{1,2}|<1$, one can further find that the annihilation of nodal lines
just corresponds to the transition from a TOTOPSC to a trivial superconductor,
indicating that here the topological structure of the normal state
plays a crucial role.

{\it Conculusion.---} We have revealed that there are two basic requirements for the realization of HOTOPSCs.
One is  the contractibility of Fermi surfaces without
crossing any TRI momentum, and the other is the presence of  nontrivial Berry phase on the
Fermi surfaces.  We have also revealed a general and simple principle to realize SOTOPSCs when the normal
state is a normal metal. Furthermore, we have shown that the realization of TOTOPSCs requires the underlying normal
state to be a topological semimetal.  Our findings can not only be applied to analyze the topological property of
intrinsic (or effective) odd-parity superconductors, but also guide us to find new promising routes to realize HOTSCs
and their concomitant Majorana modes.
In fact, we note that in a recent preprint\cite{Zhu2018mixed},  there the proposal based on a combination of  Rashba spin-orbit coupling and $s+id$ pairings
just provides an effective realization of our model in Eq.(\ref{minimal}).

Finally, it is worthy to point out that all models proposed in this work can also be taken to
describe HOTIs owing to the direct analogy between superconductors
and insulators in Hamiltonian description, in other words,  Eqs.(\ref{minimal}),
(\ref{TRI}) and (\ref{third}) are also minimal models of HOTIs.

{\it Acknowlegements.---} We would like to acknowledge
helpful discussions with Zijian Xiong, and the support
by a startup grant at Sun Yat-sen University.

\bibliography{dirac}

\end{document}